\documentstyle[12pt]{article}
\def\bA{{\bar A}}
\def\i{{\frac1{2\pi}\int_\Sigma d^2z\,}} %integral over Sigma
\def\d{\partial}
\def\na{\nabla}
\def\bd{\bar\partial}
\def\a{\alpha}
\def\b{\beta}
\def\th{\theta}

\def\g{\gamma}

\def\alg{{\bf g}}
\def\talg{{\bf * g }}
\def\gp{{\cal{G}}}
\def\lie{{\cal L}}

\def\de{\delta}
\def\P{\Phi}

\def\l{\lambda}
\def\w{\omega}
\def\om{\omega}
\def\s{\sigma}
\def\S{\Sigma}

\def\e{\epsilon}

\def\Z{\bf Z}

\def\bJ{{\bar J}}

\def\t{\tilde}

\def\ie{{\it i.e.,\/}}
\def\eg{{\it e.g.\/}}

\begin{document}

\newcommand{\inv}[1]{{#1}^{-1}} %inverse

\renewcommand{\theequation}{\thesection.\arabic{equation}}
\newcommand{\beq}{\begin{equation}}
\newcommand{\eeq}[1]{\label{#1}\end{equation}}
\newcommand{\ber}{\begin{eqnarray}}
\newcommand{\eer}[1]{\label{#1}\end{eqnarray}}
%\begin{titlepage}
\begin{center}
June, 1994
\hfill    ITP-SB-94-26\\
\hfill    RI-5-94\\
\hfill    hep-th/9406178\\
\vskip .5in
{\large \bf Introduction to Duality}\footnotemark \\
\footnotetext{To appear in ``Essays on Mirror Manifolds II''.}
\vskip .5in
{\bf Amit Giveon}\footnotemark \\
\footnotetext{e-mail address: giveon@vms.huji.ac.il}
\vskip .1in
{\em Racah Institute of Physics, The Hebrew University\\
  Jerusalem, 91904, ISRAEL} \\
\vskip .15in
and
\vskip .15in
{\bf Martin Ro\v cek}\footnotemark \\
\footnotetext{e-mail address: rocek@insti.physics.sunysb.edu}
\vskip .1in
{\em Institute for Theoretical Physics \\
State University of New York at Stony Brook \\
Stony Brook, NY 11794-3840 USA}\\
\vskip .1in
\end{center}
\vskip .4in
\begin{center} {\bf ABSTRACT } \end{center}
\begin{quotation}\noindent
We describe the duality between different geometries which arises
by considering the classical and quantum harmonic map problem.
\end{quotation}
\vfill
\eject
\setcounter{footnote}{0}
\def\baselinestretch{1.2}
\baselineskip 16 pt
\noindent
\section{Introduction and summary}
\setcounter{equation}{0}
Consider a manifold with certain geometric data: a metric, a compatible
connection with torsion, and possibly a scalar field.  Duality is a map
between different geometries; in general, it changes not
just the metric and connection on the manifold, but the topology as well.
Though there are reasons to believe that it can be generalized beyond the
cases described here, we understand how to construct the dual of a given
geometry only for geometries with isometries.

The duality map is found by considering a generalized harmonic map
problem.  Geometries that are classically dual have the same harmonic
maps. Geometries may also be dual in a stronger
sense: as explained below, they give rise to the same quantum field
theory. We call this map ``quantum duality''\footnote{When the geometry
admits $N=2$ superconformal symmetry, and if duality maps the
left-moving $N=2$ $U(1)$ current $J$ to $-J$, it is also called  a
``mirror map''.}.

We now briefly summarize our results. We begin with a manifold $M$ with a
metric $G$ (which can be written in terms of frames $e^a$ as $G=e^a\otimes
e^a$), a closed 3-form $T$, $dT=0$,\footnote{We assume that $T$ represents an
integral (possibly trivial) element of the cohomology.} and a compatible
connection $\na=d+\om$ with torsion $T^a$ that is the contraction of $e^a$ with
$T$: $T^a=e^a\rfloor T=de^a+\om^a{}_b\wedge e^b$. In a patch on $M$, the
torsion may be written as $T=\frac32dB$ for some two form $B$. We suppose also
that $M$ has some Lie group of isometries preserving $T$.  Duality with respect
to some subgroup $\gp$ of the isometry group gives a new manifold $\t M$ that
is a certain $\gp$ quotient of $\talg \times M$ (where $\talg$ is the dual of
the Lie algebra of $\gp$). On $\talg \times M$ we find the metric $\hat G$ and
the two form $\hat B$ in terms of the original metric plus two form
$E_{ij}=G_{ij}+B_{ij}$ and a basis $\{k_A\}$ of Killing vector fields
generating the Lie algebra $\alg$ of $\gp$:
\beq
\hat G +\hat B = \left(\matrix{{\t E_{AB}}&{\t E_{Aj}}\cr{}&{}\cr
{\t E_{iB}}&{\t E_{ij}}\cr}\right)=
\left(\matrix{{[E_{AB}]^{-1}}&{\t E^{AB}E_{Bj}}\cr{}&{}\cr
{-E_{iA}\t E^{AB}}&{E_{ij}-E_{iA}\t E^{AB} E_{Bj}}\cr}\right)\ \ ,
\eeq{hatE}
where
\vskip .05in
\beq
\left(\matrix{{E_{AB}}&{E_{Aj}}\cr{}&{}\cr
{E_{iB}}&{E_{ij}}\cr}\right)=
\left(\matrix{{k^i_AE_{ij}k^j_B +\l_Cf^C_{AB}}&{k^i_AE_{ij}}\cr{}&{}\cr
{E_{ij}k^j_B}&{G_{ij}+B_{ij}}\cr}\right)\ \ .
\eeq{Edef}
\vskip .05in
\noindent
Here $f^A_{BC}$ are the structure constants of $\alg$, $\l_A$ are coordinates
on $\talg$ and $X^i$ are coordinates on $M$. The dual metric $\t G$ and two
form $\t B$ are the restriction of $\hat G$ and $\hat B$ to the $\gp$
orbits of $\talg\times M$.

If the connection $\na$ has restricted holonomy, \ie\ in some subgroup of the
orthogonal group, then in general $M$ will carry some more structure.
Typically, there will be a covariantly constant $p$-form $\w_p$. If $\w_p$ is
preserved by $\gp$ as well, then duality leads to a covariantly constant $\t
\w_p$ on the dual manifold \cite{bbkim}. We give the explicit form of the dual
form when we discuss the example $\gp=U(1)$; the general case has not been
worked out.

This article is organized as follows: In the next section, we introduce
$\s$-models. In section 3, we setup the duality transformation, and in section
4, we derive the (classical) dual transformation. In section 5 we discuss
quantum duality and global issues. Finally, in section 6, we work out the
example of $\gp=U(1)$ in greater detail.

\noindent
\section{Sigma-models}
\setcounter{equation}{0}

Recall that we consider a manifold $M$ with a metric $G$, a closed 3-form
$T=\frac32dB$, and a compatible connection $\na$ with torsion $T^a$.

The generalized harmonic map problem is defined by extremizing a
functional $S$ (the ``action'') for maps $X:\S\to M$ from a Riemann
surface $\S$ (the ``worldsheet'') to the manifold $M$ (the ``target
space''). The action $S$ is:
\beq
S=-\frac1{2\pi}\int_\S \Big( ||dX||^2 vol + iX^*B\Big)\ \ ,
\eeq{maction}
where $X^*B$ is the pullback of $B$ from $M$ to $\S$ by $X$.\footnote{Since in
general $T$ is closed but need not be exact, $B$ is only defined locally, and
$-\frac{1}{2\pi}\int X^*B$ is only defined modulo $2\pi n$, where $n \in \Z$.}
Explicitly (for future use, we label this action as $S_O$),
\beq
S_O[X,\g]=-\frac1{2\pi}\int_\S d^2\s ( \sqrt\g \g^{\a\b}G_{ij}(X) +
i\e^{\a\b}B_{ij}(X))\d_\a X^i \d_\b X^j \ \ ,
\eeq{paction}
where $\g_{\a\b}$ is the metric on $\S$ (``the worldsheet metric''),
$\g=\det \g_{\a\b}$, $\a,\b=1,2$, and $i,j=1,...,D={\rm dim}\, M$.

In addition, one may couple a scalar field $\P$ (the
``dilaton'') to the curvature of the Riemann surface $\S$:
\beq
S_{dil}=\frac1{8\pi}\int_\S\P R^{(\S)}\ \ .
\eeq{dil}
Physicists call $S$ the action for a nonlinear $\s$-model; the generalized
harmonic maps are called classical solutions\footnote{When $B=\Phi=0$, these
are harmonic maps.}. As we shall describe in detail, classical duality
preserves the harmonic maps when $\S$ is a sphere.  Quantum duality is stronger
in two ways: it preserves the set of classical solutions on higher genus
Riemann surfaces as well, and it preserves the quantum field theory defined by
the functional integral $Z$ of $e^S$ with respect to $X$:
\beq
Z[\g ]=\int [DX] e^{S[X,\g ]}\ \ .
\eeq{Z}

\noindent
\section{Isometries and the duality transformation}
\setcounter{equation}{0}
Isometries are generated by Killing vectors $k_A$ where $A=1,...,d$
labels the element of the Lie algebra $\alg$ that the Killing vectors
generate:
\beq
[k_A,k_B]=f_{AB}^Ck_C \ \ .
\eeq{kill}
Here $f_{AB}^C$ are the structure constants of the Lie algebra $\alg$.
The Killing vectors preserve the metric $G$, the torsion $T$, and the
dilaton $\P$:
\ber
\lie_{k_A}G_{ij}&\equiv&\na_ik_{Aj}+\na_jk_{Ai}=0\
\ ,\nonumber\\&&\nonumber\\
\lie_{k_A}T_{ijk}&\equiv&
k_A^l\d_lT_{ijk}+\d_ik^l_AT_{ljk}+\d_jk^l_AT_{ilk}
+\d_kk^l_AT_{ijl}=0\ \ ,\nonumber\\&&\nonumber\\
\lie_{k_A}\P&\equiv&k^i_A\d_i\P=0\ \ ,
\eer{lie}
where $\lie_{k_A}$ is the Lie derivative with respect to the vector
$k_A$ (which has components $k_A^i$). When these conditions are satisfied, the
action (\ref{paction}, \ref{dil}) is invariant (modulo boundary terms)
under the transformations
\beq
\de X^i = -\e^A k_A^i
\eeq{trans}
where $\e^A$ are constant parameters.

The duality transformation is found by a two step procedure \cite{FJFT,TB,RV}:

\noindent (1) We ``gauge'' (a subgroup of) the symmetry (\ref{trans}), \ie\ we
introduce a $\gp$-connection $A$ on
$\S$, and use it to enhance the symmetry (\ref{trans}) to a ``local''
symmetry when the parameters become functions $\e^A(\s )$ on $\S$.
When the action (\ref{paction}) is invariant {\em without} boundary
terms, the procedure is called minimal coupling: We substitute
\beq
\d_\a X^i\to\na_\a X^i\equiv \d_\a X^i+A^B_\a k_B^i \ \ .
\eeq{mincoup}
More generally, one can follow the procedure of \cite{GWZW} to get a
gauge-invariant action\footnote{If a symmetry is anomalous, it is {\em not}
possible to gauge it.}.

\noindent (2) We constrain the connection $A$ to be trivial by adding a
term to the gauged action that we constructed in step (1). For classical
duality, when $\S$ is a sphere, this term is simply
\beq
S_\l[A,\l ]=\frac{i}{2\pi}\int_\S Tr\l F(A)\ \ ,
\eeq{LF}
where $F(A)=d A + A^2$ is the curvature of the connection $A$ and $\l$ is a
Lagrange multiplier in the dual $\talg$ of the Lie algebra\footnote{In certain
cases, \eg, Abelian groups, it is possible to perform a duality transformation
even when the gauging is anomalous; essentially, the anomaly can be cancelled
by a transformation of $\l$. It is sometimes possible to compactify $\l$, \eg,
in the Abelian case, for quantum duality, we take $\l$ in a toroidal subspace
of $\talg$ (see section 6).}. Extremizing the gauged action (including $S_\l$)
with respect to the Lagrange multiplier $\l$ implies that the curvature
vanishes: $F(A)=0$. On a simply connected worldsheet $\S$ such as $S^2$, this
implies that the connection $A$ is pure gauge, and in particular, can be chosen
to vanish. Thus, classically, we are dealing with the original system.  The
duality transformation is found by extremizing the gauged action (including the
Lagrange multiplier term) with respect to the connection $A$ and eliminating it
in the classical case, or more carefully, functionally integrating $A$ out in
the quantum case.

\noindent
\section{The dual action}
\setcounter{equation}{0}

For the case when we can use minimal coupling (\ref{mincoup}),
the gauged action is simply:
\beq
S[X,\g,A]=-\frac1{2\pi}\int_\S d^2\s ( \sqrt\g \g^{\a\b}G_{ij}(X) +
i\e^{\a\b}B_{ij}(X))\na_\a X^i \na_\b X^j \ \ .
\eeq{gact}
At this stage it is convenient, though by no means essential, to choose complex
coordinates on the surface $\S$. After integrating by parts and collecting
terms, the gauged action $S[X,\g,A]$ (\ref{gact}) with the Lagrange multiplier
term $S_\l[A,\l ]$ (\ref{LF}) added becomes:
\ber
S_1[X,A,\l ]=-\i \left( E_{ij}\d X^i\bd X^j + (E_{iB}\d X^i-\d\l_B)
\bA^B\right. \qquad&&\nonumber\\
\left. +A^B (E_{Bi}\bd X^i+\bd\l_B) +E_{BC}A^B\bA^C\right)\ ,&&\qquad
\eer{fact}
where the matrices $E$ are:
\beq
E\equiv \left(\matrix{{E_{AB}}&{E_{Aj}}\cr{}&{}\cr
{E_{iB}}&{E_{ij}}\cr}\right)=
\left(\matrix{{k^i_AE_{ij}k^j_B +\l_Cf^C_{AB}}&{k^i_AE_{ij}}\cr{}&{}\cr
{E_{ij}k^j_B}&{G_{ij}+B_{ij}}\cr}\right)\ \ .
\eeq{mats}
(Factors of $i$ that appear in covariant forms of various actions are absorbed
by the coordinate change $\s\to z,\bar z$.)  Extremizing the action $S_1$
(\ref{fact}) with respect to the connections $A,\bA$, leads to a new action
$S_D$ on what appears to be a larger space with coordinates $X^i,\l_A$.
However, because the first order action is invariant under the gauge
transformations
\beq
\de X^i=-\e^A(z)k^i_A\ \  ,\ \ \de\l_A=\l_C f^C_{AB}\e^B(z)\ \ ,\ \
\de A_\a^C=\d_\a\e^C+f^C_{AB}A_\a^B\e^A(z)\ \ ,
\eeq{loctra}
the dual action $S_D$ is actually defined on the orbits of the symmetry
(\ref{loctra}), which is a manifold with the same dimension $D$ that we started
with. Explicitly \cite{dOQ,GR2,GRV,AAGBL},
\ber
S_D =-\i \left( \t E_{ij}\d X^i\bd X^j + \t E_i{}^B \d X^i \bd\l_B\right.
\qquad\qquad\qquad&\nonumber\\ &\nonumber\\
\left.\qquad\qquad +\t E^B{}_i\d\l_B \bd X^i
+ \t E^{BC}\d\l_B\bd\l_C\right) \ ,\qquad&
\eer{tact}
where the dual geometry is specified by:
\beq
\hat E\equiv\left(\matrix{{\t E_{AB}}&{\t E^A{}_j}\cr{}&{}\cr
{\t E_i{}^B}&{\t
E_{ij}}\cr}\right)=
\left(\matrix{{[E_{AB}]^{-1}}&{\t E^{AB}E_{Bj}}\cr{}&{}\cr
{-E_{iA}\t E^{AB}}&{E_{ij}-E_{iA}\t E^{AB} E_{Bj}}\cr}\right)\ \ .
\eeq{tgeo}

The defining feature of duality, which clearly follows by construction and may
also be verified by explicit calculation,\footnote{See section 6.2, and ref.\
\cite{GR2}.} is that the extremal conditions that follow from $S_O$ (the
``field equations'') and the obvious condition $d^2 X^i=0$ (the ``Bianchi
identities'') are rotated into each other by the duality transformation.

\noindent
\section{Quantum duality and global issues}
\setcounter{equation}{0}
In the previous section, we discussed classical duality, namely, a
transformation between geometries that preserves their harmonic maps (from
$S^2$ to the target geometry). We now turn to quantum duality; this is a more
refined notion that preserves more structure. After a brief sketch of what a
quantum field theory (QFT) is, we consider how to preserve maps from arbitrary
surfaces $\S$ to the target geometry. This leads us to new geometries which are
orbifolds of the original geometry with respect to some subgroup (which
may be infinite or even continuous) of the isometry group \cite{RV,GR2,AAGBL}.
We then discuss the transformation of the functional measure, and show that
duality, suitably defined, is an exact symmetry of a QFT.

A quantum field theory can be defined by a set of {\em correlation functions},
associated to an action functional $S[\phi ]$ that depends on some fields
$\phi$. Formally, these may be computed from a functional integral; typically,
for some set of operators $\{ O_i(\phi )\}$, one computes
\beq
\langle \prod_i O_i\rangle = Z^{-1}\int [D\phi]e^{S[\phi ]}
\prod_i O_i(\phi )\ ,
\eeq{cor}
where
\beq
Z\equiv\langle 1\rangle=\int [D\phi]e^{S[\phi ]}
\eeq{part}
is the ``vacuum'' functional integral or ``partition function'', and
$[D\phi]e^{S[\phi ]}$ is the functional measure up to the normalization factor
$Z^{-1}$. In general, this is not well defined, but for particular cases of
interest one may make some kind of sense of these integrals\footnote{Examples
of these correlation functions are knot invariants in Chern-Simons theory
\cite{EWCS}, the Donaldson invariants on 4-manifolds \cite{don}, and
intersection forms on K\"ahler manifolds \cite{Yuk}; needless to say, not all
correlation functions are topological.}.

The partition function of the system with action $S_1[X,A,\l ]$ (\ref{fact}) is
\beq
Z=\int [DX][D\l ][DA][D\bA ] e^{S_1[X,A,\l ]}\ \ .
\eeq{part1}
We can proceed in two ways: integrating over the Lagrange
multiplier $\l$, or instead integrating over the gauge field $A$.  Although we
will eventually be able to define these integrals in such a way that the order
of integration will not matter, the issues in defining them are very different,
and we first focus on the integration over $\l$.

\noindent
\subsection{Integration over the Lagrange multiplier $\l$}
Since the Lagrange multiplier $\l$ enters the action $S_1$ (\ref{fact})
linearly, integrating over it gives a functional $\de$-function of whatever it
multiplies; this is exactly what one found in the classical case by extremizing
with respect to $\l$. When the Lagrange multiplier is in the dual Lie algebra
$\talg$, integrating over it constrains the curvature $F$ to vanish, but does
not force the connection $A$ to be trivial on a multiply connected worldsheet
$\S$.  Therefore, the theory that one gets after integrating out $\l$ (the
``F-theory'') must be integrated over all flat connections, and is not
equivalent to the original theory (the ``O-theory''). The relation between the
models is that the F-theory is the orbifold of the O-theory with respect to the
isometry group $\gp$. We can summarize this as:
\beq
[DX] \int [D\l ][DA][D\bA ] e^{S_1[X,A,\l ]} =N [DX] e^{S_F[X ]}
=N_\gp [D_\gp X] e^{S_O[X ]}\ ,
\eeq{OF1}
where $N,N_\gp$ are constant ($X$-independent) normalization factors, $S_O$ is
the action of the original model (\ref{paction}), $S_F$ is the action of the
F-theory (and formally is identical to $S_O[X]$, but is a functional of $X$
obeying different conditions), and $[D_\gp X]$ indicates that the manifold
coordinatized by $X$ is the $\gp$-orbifold of the original manifold.  If we
restrict $\l$, then we change the F-theory. For example, for $\gp=U(1)$ (see
section 6), if we choose $\l$ to be a periodic coordinate on the {\em group}
$U(1)$
(and add a boundary term to the action \cite{RV,GR2}) then the F-theory
can be made exactly equivalent to the O-theory; if one changes the periodicity
of $\l$ by a factor $k$, the F-theory becomes a $Z_k$ orbifold of the O-theory.

\noindent
\subsection{Integration over the connection $A$}
Integration over $A,\bA$ gives rise to the dual theory (the ``D-theory'') with
classical action $S_D$ (\ref{tact}).  Though the integral is gaussian, it gives
rise to factors that are not visible in the classical theory for two reasons:
(1) Being a gauge field, $A$ has a nontrivial measure, and (2) the gaussian
integral itself involves the nontrivial quadratic form $E_{AB}$ in
(\ref{mats}).  This gives rise to a functional determinant that must be
carefully defined.  This determinant has two factors: one is just a
naive factor of $[det(E_{AB})^{-1}]$ and is the correct change in
the target space volume element: $det(\t G)^{\frac12}=
det(E_{AB})^{-1}det(G)^{\frac12}$ which implies
\beq
[DX]=[det(G)^{\frac12} dX]\to [D \t X]=[det(\t G)^{\frac12} d\t X)]\ ,
\eeq{meas}
where $G$ is the metric on $M$ (the target manifold of the F-theory), $\t G$ is
the metric on $\t M$ (the target manifold of the D-theory), namely, the
metric on the $\gp$ orbits of $\talg\times M$ that follows from the classical
dual action (\ref{tact}), and $[dX],[d\t X]$ are the naive flat volume elements
on the spaces $M$, $\t M$, respectively.

The second factor arises because of hidden dependence on the worldsheet metric
$\g$ in $E_{AB}$ and in $[DA][D\bA ]$. Suitably regularized, this gives rise to
a shift in the dilaton \cite{TB}:\footnote{In some cases, there may be
further so-called higher order corrections. In the context of duality, they are
not completely understood.}
\beq
\t\Phi=\Phi + ln (det(E_{AB}))\ .
\eeq{dilshft}
We can summarize this as:
\beq
[DX][D\l ] \int [DA][D\bA ] e^{S_1[X,A,\l ]+S_{dil}[X,\g ]} =N [D\t X]
e^{S_D[\t X ]+ \t S_{dil}[\t X,\g ]}\ ,
\eeq{DF1}
where the actions $S_1,S_{dil},S_D$ are given above in
(\ref{fact}, \ref{dil}, \ref{tact}), and $\t S_{dil}$ is simply $S_{dil}$ with
$\Phi\to\t\Phi$ (\ref{dilshft}).

We emphasize that the F-theory and the D-theory are always equivalent as
quantum field theories. The question of whether duality is a symmetry between
the original O-theory and the D-theory is a matter of the global issues
discussed in the previous subsection.

\noindent
\section{An example: Abelian duality}
\setcounter{equation}{0}

In this section, we briefly discuss the case when the gauge group is Abelian
\cite{TB,RV,GR}. Specifically, we consider a target space geometry with a
$U(1)$ isometry.

\noindent
\subsection{The dual action}
Without loss of generality, away from fixed points of the $U(1)$ action on the
target space,\footnote{Fixed points lead to singularities on the dual space.}
we can choose coordinates $X=(\th,x^i)$ on $M$ where the symmetry acts by
shifts of a single periodic coordinate $\th\equiv\th+2\pi$, and the remaining
coordinates $x^i$ are left inert. In these coordinates, the background is
independent of $\th$.  The action of the original model takes the form
\ber
S_O[\th ,x]+S_{dil}=-\i \Big( E_{00}(x)\d\th\bd\th + E_{0j}(x)\d\th\bd x^j
+E_{i0}(x)\d x^i\bd\th &\nonumber\\
+E_{ij}(x)\d x^i\bd x^j-\frac14\Phi (x)R^{(\S )}
\Big)\ ,&\nonumber\\&
\eer{abel}
In the $U(1)$ case, gauging by minimal coupling is performed by the
substitution $\d\th\to\d\th +A$, $\bd\th\to\bd\th +\bA$. We also add the
Lagrange multiplier term (\ref{LF}), which here takes the form
\beq
\i ( A\bd\l-\bA\d\l )\ ;
\eeq{Adl}
up to an important total derivative, this is just $\frac{i}{2\pi}\int\l F$.
When $\l$ is chosen to have periodicity $2\pi$, as discussed in \cite{RV,GR2},
the boundary term ensures that when one integrates over $\l$, the winding modes
of $\l$ constrain the holonomy of the gauge field $A,\bA$ so that it is not
just flat, but actually trivial, and we recover the original model.

An interesting special feature of the Abelian case is that the dual model
has the same symmetry as the original model. It acts on $\l$ by
\beq
\l\to\l+\e \ ,
\eeq{abelshift}
for $\e$ constant.
If we make a duality transformation with respect to this symmetry, we
immediately see that the dual of the dual is the original model: We
gauge (\ref{abelshift}) and add a second Lagrange multiplier $\t\l$:
\ber
S_1[\th,x,A,\l ]&\to&
S[\th,x,A,\t A,\l,\t\l ]\nonumber\\
&\equiv& S_1[\th,x,A,\l ]+\i tr\Big(\t\bA A- \t A\bA +
\t A\bd\t\l-\t\bA\d\t\l\Big)\ .\nonumber\\ &&
\eer{abeldudu}
Functionally integrating out $\t A,\t\bA$ gives the constraints:
\beq
A=\d\t\l\ ,\qquad \bA=\bd\t\l \ ,
\eeq{puregauge}
and hence, after the redefinition $\th+\t\l\to\th$, we recover the original
model.  This inverse transformation is a sign of an underlying {\it group\/} of
duality transformations.  For $d$ commuting $U(1)$ symmetries, one finds an
$O(d,d,{\bf Z})$ group \cite{GR} (for a review, see, \eg,\ ref.\ \cite{GPR}).

Returning to our example and choosing a representative on the $U(1)$ orbits
(choosing ``a gauge'') such that $\th =0$, we find the gauged action
\ber
S_1[x,A,\l ]+S_{dil}=-{ \i} \Big( E_{00}A\bA+E_{0i}A\bd x^i+E_{i0}\d
x^i\bA+E_{ij}\d x^i \bd x^j && \nonumber\\
-\frac14\Phi R^{(\S )} +(A\bd\l-\bA\d\l )\Big) \ .\ \qquad &&
\hbox{\hspace{-1in}}\
\eer{abelgauge}
Functionally integrating out the gauge field $A,\bA$, we may replace the gauge
field with
\beq
A(\l ,x)\to (\d\l-\d x^iE_{i0})\inv {(E_{00})}\ ,\ \ \
\bA(\l ,x)\to -\inv {(E_{00})}(\bd\l+E_{0i}\bd x^i)\ .
\eeq{abelAbA}
Substituting (\ref{abelAbA}) into the gauged action (\ref{abelgauge}) and
recalling the shift in the dilaton that the integration over $A,\bA$ gave, we
find the dual action:
\ber
S_D[x,\l ]+\t S_{dil}=-\i\Big((\d\l-\d x^i E_{i0})\inv {(E_{00})}(\bd\l+
E_{0i}\bd x^i)\nonumber\\
+ E_{ij}\d x^i\bd x^j -\frac14(\Phi + \ln E_{00})R^{(\S )}\Big)\ .
\eer{abeldual}
{}From this we read off the dual geometry \cite{TB}:
\ber
\t G &=& \left(\matrix{\inv {(E_{00})} &\inv{(E_{00})}B_{0j}\cr & \cr
\inv{(E_{00})}B_{i0}&
G_{ij}-\inv{(E_{00})}(G_{i0}G_{0j}+B_{i0}B_{0j})\cr}\right)\ ,
\nonumber\\&&\nonumber\\&&\nonumber\\
\t B  &=&\left(\matrix{0&\inv {(E_{00})}G_{0j} \cr & \cr
-\inv {(E_{00})}G_{i0} & B_{ij}-
\inv {(E_{00})}(G_{i0}B_{0j}+B_{i0}G_{0j})\cr}\right)\ ,
\nonumber\\&&\nonumber\\
\t \Phi &=&\Phi + \ln E_{00}\ ,
\eer{dugeo}
where (recall) $E=G+B$.

\noindent
\subsection{Field Equations}
A general feature of classical duality is that field equations and Bianchi
identities are rotated into each other \cite{olddual}. Here, they are
simply interchanged.

In the original model (\ref{abel}), the field
equation and Bianchi identity are:
\beq
{\bf Field\ Equation:}\qquad\qquad\qquad\qquad \bd J+\d \bJ = 0\qquad
\qquad\qquad
\eeq{abelfeq}
for
\beq
J=E_{00}\d\th+E_{i0}\d x^i\ ,\qquad \bJ=E_{00}\bd\th+E_{0i}\bd x^i \ ,
\eeq{abelJbJ}
and
\beq
{\bf Bianchi\ Identity:}\qquad \d[\inv {(E_{00})}(\bJ-E_{0i}\bd x^i)]-\bd
[(J-\d x^i E_{i0})\inv {(E_{00})}]=0
\eeq{abelBi}
(substituting the definition of $J,\bJ$ (\ref{abelJbJ}) into
(\ref{abelBi}), this becomes just the triviality $\d\bd\th-\bd\d\th=0$).
We may write these as:
\beq
d*J=0\ ,\ \ dA= 0\ ,
\eeq{djda}
for $A=d\th$.

In the dual model, we have the obvious currents
\beq
\t J=\d\l \ ,\qquad \t{\bJ}=-\bd\l \ .
\eeq{abeltJ}
These obey the dual Bianchi identity:
\beq
{\bf Dual\ Bianchi\ Identity:}\qquad\qquad\qquad\bd\t J +\d\t{\bJ}=0
\ .\qquad\qquad
\eeq{abelduBi}
By construction, the $\l$ field equation is $F(A,\bA )=0$, with
$A,\bA$ given in (\ref{abelAbA}); in terms of the dual currents $\t
J,\t{\bJ}$, this takes the form:
\beq
{\bf Dual\,\, Field\,\, Equation:}\ \d[\inv {(E_{00})}(\t{\bJ}-E_{0i}\bd
x^i)]-\bd [(\t J-\d x^i E_{i0})\inv {(E_{00})}]=0
\eeq{abeldufeq}
In the compact notation of (\ref{djda}), these become:
\beq
d*\t J = 0\ ,\ \ dA= 0\ ,
\eeq{djtda}
where now $\t J = *d\l$ and $A=A(\l ,x)$. Thus we see that the field equation
(\ref{abelfeq}) becomes the dual Bianchi identity (\ref{abelduBi}) and the
Bianchi identity (\ref{abelBi}) becomes the dual field equation
(\ref{abeldufeq}).

\noindent
\subsection{Further structures}
When the original manifold has restricted holonomy, it can have a covariantly
constant $p$-form $\w_p$.  Very recently, it was shown by B.\ B.\ Kim
\cite{bbkim} that if the $p$-form is independent of $\th$ (or, in invariant
language, if $\lie_k \w_p = 0$), then one can find a dual covariantly
constant form $\tilde{\om}$ on the dual space $\t M$:
\beq
\tilde{\om}_{0 k_{1}\cdots k_{p-1}}=\inv{(E_{00})}\om_{0 k_{1}\cdots
             k_{p-1}}
\eeq{om1}
\beq
\tilde{\om}_{jk_{1}\cdots k_{p-1}}=\om_{jk_{1}\cdots k_{p-1}}
            -\inv{(E_{00})}W_{jk_{1}\cdots k_{p-1}}\ ,
\eeq{om2}
where
\ber
W_{jk_{1}\cdots k_{p-1}} &=& E_{j0}\om_{0k_{1}\cdots k_{p-1}}
-E_{k_{1}0}\om_{0jk_{2}\cdots k_{p-1}} \nonumber \\ & &
+E_{k_{2}0}\om_{0k_{1}jk_{3}\cdots k_{p-1}} \nonumber \\ & & \cdots \nonumber
\\ & & +(-1)^{p-1}E_{k_{p-1}0}\om_{0k_{1}\cdots k_{p-2}j}\ .
\eer{omw}
Thus the dual connection (in general {\em with} torsion) also has restricted
holonomy.

\vskip 1.0in

\noindent
{\bf Acknowledgements}

\bigskip

\noindent
We thank Itai Hareven for his comments on the manuscript.
This work is supported in part by BSF-American-Israeli Bi-National Science
Foundation. AG thanks the ITP at Stony Brook and MR thanks the Racah Institute
for their respective hospitality. The work of AG is supported in part by an
Alon Fellowship. The work of MR is supported in part by NSF Grant No.\ PHY 93
09888.

\end{document}